\documentclass[twocolumn,prl,aps,epsf]{revtex4}

\usepackage{graphicx}
\usepackage{dcolumn}
\usepackage{bm}
\newcommand{\be}{\begin{equation}}
\newcommand{\ee}{\end{equation}}
\usepackage{wasysym}
\usepackage{amssymb}

\begin{document}

\title{Topological Entanglement Entropy of a Bose-Hubbard Spin Liquid}

\author{Sergei V. Isakov}
\affiliation{Theoretische Physik, ETH Zurich, 8093 Zurich, Switzerland }
\author{Matthew B. Hastings}
\affiliation{Duke University, Department of Physics, Durham, NC, 27708}
\affiliation{Microsoft Research, Station Q, CNSI Building, University of California, Santa Barbara, CA, 93106}
\author{Roger G. Melko}
\affiliation{Department of Physics and Astronomy, University of Waterloo, Ontario, N2L 3G1, Canada}

\date{\today}

\maketitle

{\bf
The Landau paradigm of classifying phases by broken symmetries was demonstrated to be incomplete when it was realized that
different quantum Hall states could only be distinguished by more subtle, topological properties \cite{WenTO}.
Today, the role of topology as an underlying
description of order has branched out to include topological band insulators,
and certain featureless gapped Mott insulators with a topological degeneracy in the groundstate wavefunction.  
Despite intense focus, very few  candidates for these topologically ordered ``spin liquids'' exist.

The main difficulty in finding systems that harbour spin liquid states is the very fact
that they violate the Landau paradigm, making conventional order parameters non-existent.
Here, we uncover a spin liquid phase in a 
Bose-Hubbard model on the kagome lattice, and measure its topological order directly via the {\emph {topological entanglement entropy}}.  This is the first smoking-gun demonstration of a non-trivial spin liquid, identified through its entanglement entropy as a gapped groundstate with emergent $Z_2$ gauge symmetry.
}

Quantum spin liquid phases are notoriously elusive, both in experimental materials and in theoretical models.
In part, this is due to the delicate balance of microscopic interaction that must occur so that conventional 
symmetry-broken order is suppressed to low temperatures.  The search is also hampered by the lack
of a measurable order parameter, such as magnetization, that would offer a positive indicator of spin liquid 
behavior.  Instead, the current procedure of identifying spin liquids involves eliminating all possible order parameters
through exhaustive
searches of correlation functions \cite{Meng}.
Theoretical work has nonetheless developed a classification scheme of gapped spin liquid states 
based on the {\it topological} degeneracy of their wavefunction.
In fact, there exist several model Hamiltonians that have been proposed to contain 
spin liquid groundstates with the most trivial $Z_2$ topological order (corresponding to a four-fold degeneracy on a torus).
One of the earliest was the triangular-lattice quantum dimer model \cite{Roderich} -- where
dimers are intended to be an effective description of local singlet correlations, not physical spins.
Another paradigm in studying topological order has recently emerged in the 
toric code \cite{Kitaev}, as it is the simplest of a class of exactly solvable models (such as the Levin-Wen models \cite{LWmodels}) describing
different topological quantum field theories.
Unfortunately, these models require somewhat artificial multi-spin interaction terms.

\begin{figure} {
\includegraphics[width=3.2 in]{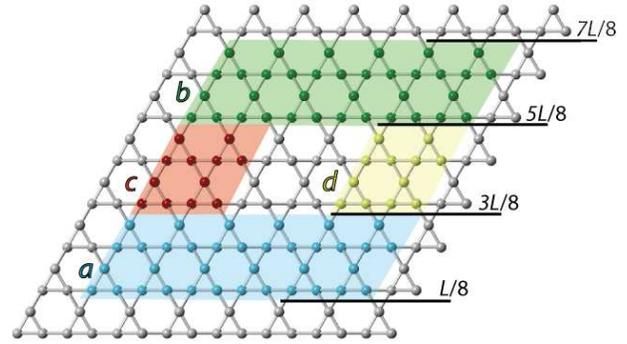} \caption{ An $L=8$ toroid (a periodic lattice with $8 \times 8 \times 3$ spins).  In Equation~\ref{Gamma}, the subregions are
$A_1 = a \cup b \cup c \cup d$, $A_2 = a \cup b \cup c$, $A_3 = a \cup b \cup d$, and $A_4 = a \cup b$.  The width of the annulus is $R=6$, and its thickness is $r=2$.
\label{lattice}
}
} \end{figure}

In order for a model to be relevant for real physical systems,
it is essential to find a spin Hamiltonian with simple two-body interaction terms
and a topologically ordered spin liquid state.
One important step in this direction occurred with the introduction of 
a kagome Bose-Hubbard model by Balents, Girvin and Fisher \cite{Leon1}.
Although this model has a four-spin interaction, it 
contains a $Z_2$ spin liquid over an extended region of its phase
diagram.  Previously, a related model was shown through quantum Monte Carlo (QMC) studies to have a featureless Mott insulating 
state \cite{sergeiSL,cooppara}, where, like most experimental candidates, the absence of order in correlation functions 
was the main evidence for spin liquid behavior.

In 2006, Levin and Wen \cite{LW} and Kitaev and Preskill \cite{KP} identified a quantity $2\gamma$ called the topological entanglement entropy (EE) which
is designed to replace the concept of an ``order parameter'' in a topologically ordered system.  Based on the idea that the spin liquid state
is a type of collective paramagnet, the topological EE is designed to pick up non-local correlations in the groundstate wavefunction that are not 
manifest as conventional long-range order.  However, these correlations contribute to the total entanglement between different subregions
of the system $A$ and its complement $B$ (where $A \cup B$ is the entire system).  The EE between $A$ and $B$ can be quantified by
the Renyi entropies,
\be
S_n(A)=\frac{1}{1-n} \ln \left[{ {\rm Tr}(\rho_A^n) }\right],
\ee
where $\rho_A$ is the reduced density matrix of region $A$.  In a topologically ordered state, the 
non-local entanglement gives a topology-dependent subleading correction to ``area-law'' scaling of the EE of subregion
$A$.  In 2D, $S_n(A)= a \ell - \gamma j + \cdots$ where $a$ is a non-universal constant, $\ell$ is the boundary length between $A$ and $B$, and
$j$ is the number of disconnected boundary curves.  
In Levin and Wen's \cite{LW} construction (used in the calculations in this paper), the 
topological contribution can be isolated from the area-law scaling (plus any corner contributions) by considering separately the
Renyi \cite{PI} entropies on four differently shaped subregions (Figure~\ref{lattice}),
\begin{equation}
2\gamma = \lim_{r,R \rightarrow \infty} \left[{ -S_n(A_1) + S_n(A_2) + S_n(A_3) - S_n(A_4) }\right]. \label{Gamma}
\end{equation}

Naively, since calculating $\gamma$ requires complete knowledge of the groundstate wavefunction (through $\rho_A$), previous efforts to calculate it have been restricted to models which can be solved exactly either analytically (e.g. the toric code) or through numerical exact diagonalization on small size systems (e.g.~the triangular lattice dimer model \cite{TriangTEE}).  The ability to use $\gamma$ as a general tool to search for and characterize non-trivial topologically ordered phases has been hindered by the inability to access the wavefunction in large-scale numerical methods, namely QMC, currently the only
scalable quantum simulation method in 2D and higher.  However, with the recent introduction of measurement methods based on the {\it replica trick},
QMC is now able to access $S_n(A)$ for $n \geq 2$ \cite{XXZ}, therefore giving one a method to calculate $\gamma$ in large-scale simulations
of quantum spin liquids.

\begin{figure*} {
\includegraphics[width=6in]{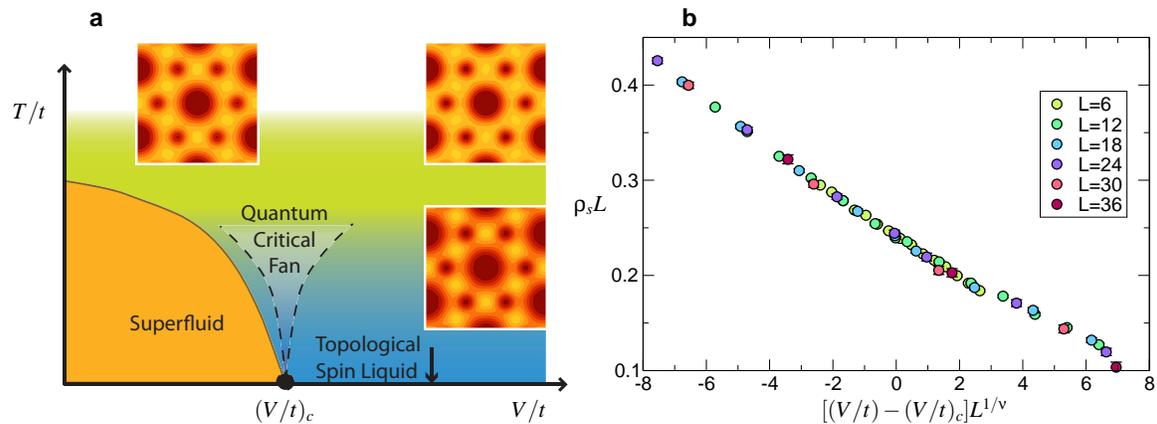}
\caption{ {\bf (a)} The schematic phase diagram of the kagome Bose-Hubbard model (\ref{ham}).  Insets are the structure factor (Fourier transform of the equal-time density-density correlation function) for $V/t=6$ and 8 at high $T$ ($\beta=3$, top) and low $T$ ($V/t=8$ and $\beta=48$, bottom) in the spin liquid phase.
{\bf (b)} Data collapse of the superfluid density, 
which in the vicinity of a continuous phase transition should scale as
$
 \rho_s(L,V/t,\beta)=L^{-1}F([V/t-(V/t)_c]L^{1/\nu},\beta/L^z).
$
Here, $F$ is the scaling function, $z$ is the dynamical critical exponent, and $\nu$ is the correlation length exponent. It follows from the above equation that if we plot $\rho_sL$ as a function of $[V/t-(V/t)_c]L^{1/\nu}$ at fixed $\beta/L^z$ then the curves for different system sizes should collapse onto the universal curve $F$ for appropriate values of $\nu$ and $(V/t)_c$, as  shown
for $\nu=0.6717$, $(V/t)_c=7.0665$, and $\beta/L=2$.
\label{fig:PhaseD}
}
} \end{figure*}

\begin{figure*}
{
\includegraphics[width=6.5in]{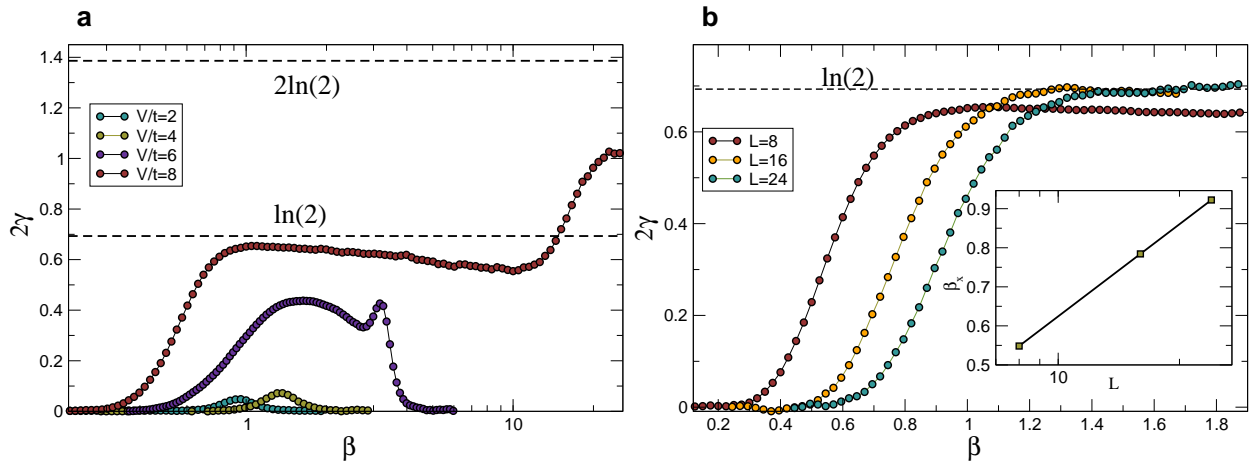}
\caption{ {\bf (a)} The topological entanglement entropy (EE) measured on an $L=8$ system as a function of inverse temperature $\beta = t/T$.  The plateaus are a measure of the total quantum dimension \cite{LW}, and should be $\ln(2)$ and $2 \ln(2)$ for a $Z_2$ spin liquid.  {\bf (b)} The approach of the topological EE to the first plateau, for different system sizes.  The value of the crossover temperature ($\beta_{\rm x}$), measured at $2 \gamma = \ln(2)/2$, shows a logarithmic dependence on system size (inset).
\label{fig:tent}
}
} \end{figure*}

Using Stochastic Series Expansion QMC \cite{SSE,SSE1}, we simulate a hard-core Bose-Hubbard model on the kagome lattice, with nearest-neighbor hopping and a six-site potential around each lattice hexagon,
\be
H = -t \sum_{\langle ij \rangle} [b^{\dagger}_i b_j + b_i b^{\dagger}_j] + V \sum_{\hexagon} (n_{\hexagon})^2, \label{ham}
\ee
where $b^{\dagger}_i$ ($b_i$) is the boson creation (annihilation) operator, and $n_{\hexagon} = \sum_{i \in \hexagon} n_i$, where $n_i=b^{\dagger}_ib_i$ is the number operator.
As mentioned above, variations of this model with more complicated spin interactions are known to harbour a robust spin liquid groundstate \cite{Leon1,sergeiSL,cooppara}.  In this paper, we consider the simplified Hamiltonian (\ref{ham}),
with only nearest-neighbor hopping, which may be more amenable to construction for example in real cold atomic systems.
We observe a transition at low temperature between a superfluid phase and an insulating phase for $(V/t)_c \approx 7.0665(15)$
(Figure \ref{fig:PhaseD}).
For $V/t > (V/T)_c$ the superfluid density scales to zero, and density and bond correlators are featureless (similar to Ref.~\onlinecite{sergeiSL}).
This strongly suggests that the insulating phase is a spin liquid. 
To characterize it, we calculate the topological EE, Equation~(\ref{Gamma}) with $n=2$, which 
 for a $Z_2$ topological phase should
approach $2 \ln(2)$ in the limit $T \rightarrow 0$ \cite{LW}.
The regions $A_i$ are shown in Figure \ref{lattice} for an $L=8$ system; these are scaled proportionally for the other systems sizes studied in this paper, where $L$ is always a multiple of 8.  Results for $\gamma$ as a function of inverse temperature $\beta = t/T$ are
shown in Figure \ref{fig:tent} for several $V/t$.

In the topological phase ($V/t=8$) we see two distinct plateaus, at differing temperatures, with a non-zero topological EE as $T\rightarrow 0$.  The phenomenon is known to occur in other models such as the toric code \cite{castelnovo}, where the topological EE at zero
temperature of $2\ln(2)$ can be viewed as a sum of
electric and magnetic contributions, each contributing $\ln(2)$.  If the electric and magnetic defects have different energies, theory predicts two distinct plateaus corresponding to these individual crossover temperatures \cite{castelnovo}, as seen in our data.  
However, at any fixed non-zero temperature, in the limit of large $L$, the topological EE vanishes, as the
probability of having thermally excited defects in the annulus $A_1$ (Figure \ref{lattice}) tends to unity.  Indeed, under the assumption that the probability of having a defect is proportional to $L^2 \exp(-E/k_{\rm B} T)$, where $E$ is the defect energy,
the temperature required to see accurate plateaus in the topological EE scales logarithmically with
$L$.  In Figure \ref{fig:tent}, we show finite-size scaling data consistent with this logarithmic scaling. 
Note that our value for the topological EE at the higher-$T$ plateau is indeed very close to $\ln(2)$, and becomes more accurately quantized
at larger system sizes.  The value of $2\ln(2)$ at the lower-$T$ plateau for $L=8$ is not as accurately quantized, but is still approached.

In the superfluid phase, the topological EE tends to zero as $T\rightarrow 0$ (Figure \ref{fig:tent}).  However, surprisingly, for
$V/t=6$
we observe
a plateau in the topological EE at intermediate temperatures, $T\sim t$.  One possible explanation for this
plateau can be understood by thinking of a simpler phase transition present in the toric code, induced by adding
a parallel magnetic field.  Consider a square-lattice toric code Hamiltonian $H=-U \sum_{+}\prod_{i\in +} S^z_i - g \sum_{\Box}\prod_{i\in \Box} 
S^x_i - h \sum_i S^z_i$, where the first vertex term term penalizes vertices that do not have an even number of up spins on the legs of the neighboring bonds, and the second sum is over plaquettes.  Suppose $U\gg g$.  By increasing $h/g$, we induce a phase transition from a topological phase to a trivial phase.  In a non-zero temperature regime where $U\gg T \gg g$, the problem becomes classical: the quantum dynamics induced by $g$ can be ignored and the
vertex term restricts us to states described by closed loops of up spins.  In this classical problem, there is a phase transition as $h/T$ increases from a topological phase with long loops to a trivial phase with only short loops (this transition is dual to the
2D Ising transition).  Thus, at high temperatures (small $h/T$) we see a topological EE, while at lower $T$ the
topological EE disappears.  In fact, this kind of physics has been suggested to occur as a ``cooperative paramagnet" in a related
kagome lattice model \cite{cooppara}.

The QMC results indicate a quantum critical point, separating the superfluid and topological phases, for a critical $V/t$ located precisely by studying the finite size scaling of the superfluid density $\rho_s$ (Figure \ref{fig:PhaseD}).
The data scales very well with the dynamical exponent $z=1$ and the XY value \cite{xyexp} for the correlation length exponent $\nu=0.6717$ at $(V/t)_c=7.0665(15)$.
At $T=0$ near the quantum critical point, we
expect good quantization of the topological EE whenever $L$ is sufficiently large compared to the correlation length $\xi$, so
the topological EE may be controlled by a scaling function of $L/\xi$.  At $T>0$, the scaling of the topological EE in the
quantum critical fan appears not to have been considered previously; the plateau at intermediate temperatures for $V/t=6$ can perhaps also be
understood as a manifestation of increasing $T$ moving one from the zero temperature trivial phase into the quantum critical fan.  Even
at $T=0$, the behavior of constant terms in the entropy at a critical point is largely unexplored, and may depend sensitively upon the
geometry used to define it \footnote{For example, in the annulus geometry used here, each term in the entropy is predicted to have logarithmically
divergent corner  corrections \cite{logcorner,ryu} at a critical point.  While the divergent portion of the corner terms cancels out in the
difference we consider, a constant piece could be left over from the corners at the critical point due to long-range correlations.  If instead we extract
a constant term from a cylindrical geometry without corners as in \cite{Max} we might find a different result}.  Scaling predicts that
near the critical point the topological EE is a function of $\beta/\xi$ and $L/\xi$, implying that in the topological phase, corrections to $2\ln(2)$ should depend upon $(L/\xi)^2 \exp(-\beta/\xi)$, consistent with a defect energy of order
$1/\xi$.  Based on the intermediate temperature plateau at $V/t=6$, it seems likely that only one type of defect, the magnetic defect, becomes gapless at criticality.

{\it Discussion}  - In order to identify a topological phase, it is essential to perform non-local probes.  Experimentally, such
non-local probes could involve braiding operations as in the proposal of \cite{dassarma}.  In this paper, we have
shown that the topological entanglement entropy (EE), calculated by
QMC using the replica trick, is a practical numerical non-local probe.  Other probes might be possible,  such as calculating the ground state
degeneracy on lattices of different topology.  However, that probe suffers from two drawbacks.  First, studying surfaces of
different Euler characteristic requires
introducing defects, which is undesirable \footnote{One interesting possibility is studying a system on the Klein bottle, which also has Euler characteristic zero and hence does not require defects in the lattice.  The $Z_2$ theories have the same degeneracies on the Klein
bottle but other theories may have different degeneracy.}.  Second, while it would be possible in QMC to calculate the ground state degeneracy
by integrating the specific heat, it requires accurate simulations at a temperature low enough to suppress all excitations - a regime where simulation ergodicity typically becomes a problem.
In contrast,  we have demonstrated that measurement of topological EE yields accurate quantization once one has suppressed all topologically non-trivial
defects -- and indeed one even sees accurate quantization at higher temperature where only one kind of defect is suppressed.  
Thus, we expect that replica QMC measurements of topological EE will be  a fundamental technique in the characterization of non-trivial topological phases in the future.

This work has been supported by the Natural Sciences and Engineering Research Council of Canada (NSERC), NSF grant No PHY 05-51164 (KITP) and the Swiss HP$^2$C initiative. Simulations were performed on the Brutus cluster at ETH Zurich and the computing facilities of SHARCNET.

%\bibliographystyle{naturemag}
%\bibliography{Biblio}

\end{document}